\documentclass[superscriptaddress,nofootinbib,twocolumn]{revtex4}
\usepackage{amsmath,lscape,epsfig}
\usepackage{amsfonts}
\usepackage{amsmath}
\usepackage{amssymb}
\usepackage{slashed}
\usepackage[utf8]{inputenc}
\usepackage{graphicx}
\usepackage{epstopdf}

\def\beq{\begin{equation}}
\def\eeq{\end{equation}}
\def\beqa{\begin{eqnarray}}
\def\eeqa{\end{eqnarray}}
\def\ban{\begin{eqnarray*}}
\def\ean{\end{eqnarray*}}
\def\bi{\begin{itemize}}
\def\ei{\end{itemize}}

\begin{document}

\title{Properties of magnetized neutral mesons within a full RPA evaluation}

\author{Sidney S. Avancini} \email{sidney.avancini@ufsc.br}
\affiliation{Departamento de F\'{\i}sica, Universidade Federal de Santa
  Catarina, 88040-900 Florian\'{o}polis, Santa Catarina, Brazil}

\author{William R. Tavares} \email{williamr.tavares@hotmail.br}
\affiliation{Departamento de F\'{\i}sica, Universidade Federal de Santa
  Catarina, 88040-900 Florian\'{o}polis, Santa Catarina, Brazil}

\author{Marcus B. Pinto} \email{marcus.benghi@ufsc.br}
\affiliation{Departamento de F\'{\i}sica, Universidade Federal de Santa
  Catarina, 88040-900 Florian\'{o}polis, Santa Catarina, Brazil}

\begin{abstract}

We consider the two flavor Nambu--Jona-Lasinio model within the RPA framework to evaluate the masses of 
the $\sigma$ and $\pi^0$ mesons and the $\pi^0$ decay constant  in the presence of a magnetic field at
vanishing temperatures and baryonic densities. The present work extends  other RPA applications by fully 
considering the external momenta which enter   the  integrals  representing  the magnetized polarization 
tensor. We employ a  a field independent regularization scheme so that more accurate results can be obtained 
in the  evaluation  of physical quantities containing 
pionic contributions. As we show, this technical improvement generates results which agree well with those 
produced by lattice simulations and chiral perturbation theory.  Our  method      may also proves
to be useful in future evaluations of  quantities such as the shear viscosity and  the equation of state of 
magnetized quark matter with mesonic contributions.

\end{abstract}

\maketitle

\vspace{0.50cm}
PACS number(s): {12.38.Lg, 12.39.-x, 12.38.-t, 13.40.-f }
\vspace{0.50cm}

\section{Introduction}

The study of strongly interacting magnetized matter has been receiving much attention recently due to the fact
that this type of matter may be produced in peripheral heavy ion collisions \cite {kharzeev}  apart from 
possibly being present in magnetars \cite {magnetars}. In both situations the  magnitude of the magnetic 
fields is huge and may reach about
$\sim 10^{19} \,G$ and $\sim 10^{18} \,G$ in each case respectively. As far as heavy ion-collisions are
concerned the presence of a strong magnetic field
most certainly plays a role  despite the fact that,   in principle,
the   field intensity should decrease very rapidly  
lasting for about 1-2 fm/c only \cite {kharzeev}. The possibility that
this short time interval may    
\cite{tuchin} or may not \cite{mclerran} be affected by conductivity
remains under dispute. Current theoretical investigations analyze, eg, the influence of the magnetic 
field in the quantum chromodynamics (QCD) phase diagram and related quantities such as the order parameters 
for the chiral and deconfinement transitions as well as the equation of state to be used in stellar 
modelling. To perform evaluations one usually employs lattice QCD or model approximations. The first 
approach is limited by the notorious sign problem which prevents its application at finite densities. 
Within the second approach one usually considers some effective theory, such as the Nambu--Jona-Lasinio model (NJL),
in the framework of a given approximation such as the traditional mean field approximation (MFA) which is
used in most applications. A summary of  results covering these topics can be found in recent 
reviews \cite {reviews}. On the other hand, the analysis of other important quantities such as those 
related to mesonic modes has received less attention in the literature despite their 
phenomenological importance. It is important to note that even the properties of neutral  mesons
may be affected by the external magnetic field produced in the earliest phase of heavy-ion collisions 
due to their quark content.  Usually, the evaluation of mesonic observables, such as  masses and 
decay constants, is more cumbersome than the MFA approximation evaluation of the equation of state (EoS) 
for example. The reason is that the momentum flow within the integrals representing quarkionic loops 
may depend on the external momentum, $k$, so that the sum over Matsubara's frequencies and/or Landau levels
become more cumbersome. One could then use a further approximation by considering $I(k^2) \approx I(0)$ 
which is the approach considered in an 
 early evaluation of mesonic properties in the presence of a background magnetic field \cite {lemmer} 
 where the authors have considered the NJL model in the RPA framework performing  the evaluations with 
 Schwinger's proper time approach. The same model has been recently considered within the Ritus 
 formalism in Ref. \cite {sadooghi}   where  the authors have adopted the lowest Landau level (LLL)
approximation in order to regularize their integrals. However, this procedure  generates unwanted 
tachyonic instabilities at low temperatures and strong magnetic fields. Another powerful tool which
has been considered in the evaluation of pionic observables is chiral perturbation theory (ChPT) 
\cite {andersen}. Finally, we point out that the mass of magnetized pions  has also been evaluated using lattice QCD 
(LQCD) simulations  \cite {hidaka}.
In the present work we use the RPA, within a magnetic field independent regularization scheme (MFIR)  \cite {klimenko,nosso1}, 
to fully evaluate the magnetized polarization tensor without 
any further approximations in order to describe  the phenomenology of magnetized pions in a more 
accurate fashion. To the best of our knowledge this technically non trivial task, which requires 
the  sum over  Landau levels contributing to  momentum dependent loops, has not been carried out before.     
Here, we employ the  (full) RPA  method to evaluate the $\sigma$ meson  mass ($m_\sigma$) as well 
as the $\pi^0$ meson mass ($m_{\pi^0}$), decay constant ($f_{\pi^0}$), and its coupling to 
quarks ($g_{\pi^0 q}$). 
In particular, our numerical results for $m_{\pi^0}$  and   $f_{\pi^0}$ agree well, respectively, 
with LQCD and ChPT predictions. It is important to remark  that by establishing a reliable technique 
to evaluate the magnetized polarization tensor our work also represents an important step towards
the complete evaluation of pionic contributions to the equation of state describing magnetized 
quark matter \cite {fukushima} as well as shear viscosity \cite {lang} among other possible 
future applications.
The pseudo-scalar polarization loop was also treated in the RPA approximation in Ref. \cite {fukushima} using a similar
approach to the one adopted in the present paper.  Since the motivation of the  
former work was to propose a mechanism to explain 
the inverse magnetic catalysis, no systematic study of neutral mesons properties under strong magnetic fields was performed.
In fact, one of the major goals of that study  was to analyze the role of the $\pi^0$ in producing an effect opposed to 
the magnetic catalysis.
Then, a rather complex expression for the pion propagator was written down  in terms of Laguerre polynomials and integrals which 
would certainly be inconvenient for further generalizations. Also, an approximate version of the pion propagator
as an expansion for low momentum was obtained. Next, in order to justify the performed approximations, the full expression 
was evaluated numerically confirming the accuracy of the approximated expression. Here, on the other hand, one of our main contributions
is to provide a simple, and exact, analytical expression for the polarization tensor which, although similar to Eq.(10) 
of Ref. \cite {fukushima},  can be used without any restriction. 
As Ref. \cite {fukushima} shows, understanding the hadron properties under strong magnetic fields 
is important to understand the QCD phase diagram. In this sense, the present work adds to the theoretical effort by 
providing an elegant and accurate  analytical treatment of the magnetized polarization tensor.
The work is organized as follows. In the next section we present the two flavor NJL model in 
the presence of a magnetic field and evaluate the neutral pion mass and decay constant. In Sec. III 
we compare  our numerical results with ChPT and LQCD. Our conclusions are presented in Sec IV. 
For completeness, an appendix containing technical details is also included.


\section{General formalism}
\noindent In the presence of an electromagnetic field the two-flavor NJL model can be described by
\begin{multline}
\mathcal{L}=\overline{\psi}\left(i \slashed D - \tilde{m}\right)\psi
+G\left[(\overline{\psi}\psi)^{2}+(\overline{\psi}i\gamma_{5}\vec{\tau}\psi)^{2}\right]-
\frac{1}{4}F^{\mu\nu}F_{\mu\nu} ~,
\end{multline}
where $A^\mu$, $F^{\mu\nu} = \partial^\mu A^\nu - \partial^\nu A^\mu$ 
are respectively the electromagnetic gauge and  tensor fields, $G$ represents the coupling 
constant, $\vec{\tau}$ are isospin Pauli matrices,  $Q$ is the diagonal quark charge 
\footnote{Our results are expressed in Gaussian natural units 
where $1\,{\rm GeV}^2= 1.44 \times 10^{19} \, G$ and $e=1/\sqrt{137}$.} matrix, 
Q=diag($q_u$= $2 e/3$, $q_d$=-$e/3$),
$D^\mu =(i\partial^{\mu}-QA^{\mu})$ is the covariant derivative,   
 $\psi$ is the quark fermion field, and $\tilde{m}$ represents the bare quark mass matrix,
\begin{equation}
 \psi = \left(
\begin{array}{c}
\psi_u  \\
\psi_d \\
\end{array} \right) ~, ~
 \tilde{m}  = \left(
\begin{array}{cc}
m_u & 0 \\
0 & m_d \\
\end{array} \right) ~.
\end{equation}
In the present work we  consider the NJL model within the mean field approximation (MFA), which 
is obtained through a
linearization of the  $\mathcal{L}$ interaction term disregarding quadratic fluctuations. Taking into 
account that the pseudo-scalar condensate vanishes 
due to parity considerations, one obtains \cite{buballa}: 
\begin{equation}
 \mathcal{L}=\overline{\psi}\left(i\slashed D-M\right)\psi+G \left \langle \overline{\psi}\psi \right \rangle^{2}-
 \frac{1}{4}F^{\mu\nu}F_{\mu\nu}~,
\end{equation}
\noindent where the constituent quark mass is defined by 
\begin{equation}
 M=m-2G \left \langle \overline{\psi}\psi \right \rangle. \label{gap}
\end{equation}
We consider here $m=m_u$=$m_d$  and choose the Landau gauge, $A^{\mu}=\delta_{\mu 2}x_{1}B$, which satisfies 
$\nabla \cdot \vec{A}=0$ and $\nabla \times \vec{A}=\vec{B}=B{\hat{e_{3}}}$
, i. e., resulting in a constant magnetic field in the z-direction.

\begin{figure}[h]
\begin{tabular}{ccc}
\includegraphics[width=7.cm]{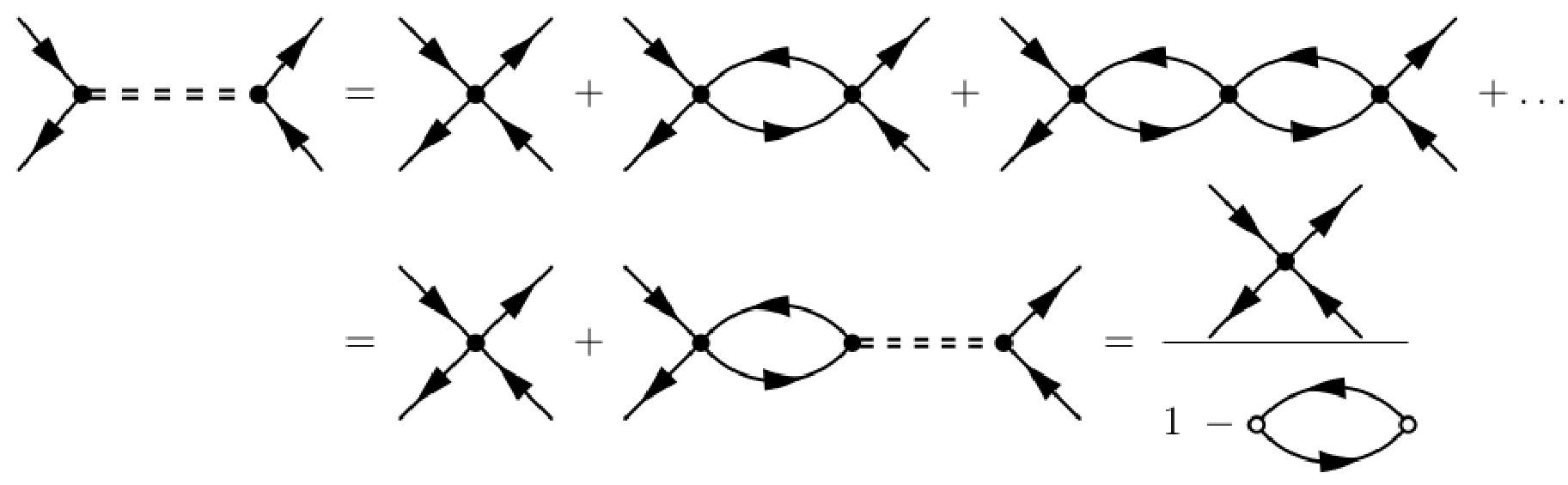}\\
\end{tabular}
\caption{Diagrammatic representation of the RPA approximation. }
\label{RPA}
\end{figure}

The $\pi^0$ pole mass calculation in the presence of  strong magnetic fields
 is performed generalizing the 
derivation of Refs. \cite{NJL-122,kleva}.  The 
pion and sigma mesons  are associated respectively to scalar and pseudoscalar collective modes. 
In the following, in order to point out the new features that arise because of the presence 
of the external magnetic field, we briefly discuss the main steps involved in  the $\pi^0$ mass calculation
within the two flavor NJL model.
The $T$-matrix for the scattering of pairs of quarks, $(q_1q_2) \rightarrow (q_1^\prime q_2^\prime)$, can be 
calculated by solving the Bethe-Salpeter equation in the ladder or random phase approximation (RPA). 
Being mainly interested in pionic degrees of freedom  
we start by considering the pseudoscalar channel within the RPA which  formally consists
in summing the
geometric series diagrammatically represented in Fig. \ref{RPA}. The left hand side of the equality in 
Fig. \ref{RPA} 
can be  calculated by representing  the quark-pion interaction with the following term \cite{kleva}:
\begin{equation}
 \mathcal{L}_{\pi qq}=i g_{\pi qq} \overline{\psi} \gamma_5 \vec{\tau}\cdot \vec{\pi} \psi~,
\end{equation}
\noindent where  $\vec{\pi}$ stands for the pion field while $g_{\pi qq}$ represents the coupling strength
between pions and quarks. The right hand side of Fig. \ref{RPA} is calculated using the NJL model and selecting
the quantum numbers associated to the neutral pion, $\pi^0$. Then,  
one can easily show that the effective interaction is given by the relation:
\begin{equation}
  (ig_{\pi^{0} qq})^2~iD_{{\pi}^0}(k^2)=\frac{2iG}{1-2G\Pi_{ps}(k^{2})} ~, \label{BS}
\end{equation}
where the pseudo-scalar polarization loop reads:
\begin{equation}
\frac{1}{i}\Pi_{ps}(k^{2})=-\sum_{q=u,d} \int \frac{d^{4}p}{(2\pi)^{4}}
Tr[i\gamma_{5}iS_q (p+\frac{k}{2})
i\gamma_{5} i S_q (p-\frac{k}{2})].  \label{loop-ps}
\end{equation}
Proceeding in an analogous way one obtains 
\begin{equation}
\frac{1}{i}\Pi_{s}(k^{2})=-\sum_{q=u,d} \int \frac{d^{4}p}{(2\pi)^{4}}
Tr[iS_q(p+\frac{k}{2})iS_q(p-\frac{k}{2})]~, 
\label{loop-sc}
\end{equation}
for the scalar channel.
In the evaluation of the polarization loops one needs  the  propagators representing  
mesons, $D_{{\pi}^0}(k^2)$, as well as  quarks, $S_q(k^2)$, as we  discuss next.
\begin{figure}[h]
\begin{tabular}{ccc}
\end{tabular}
\end{figure}
%
%
%
%

\noindent Since $\pi^0$ is uncharged, $D_{\pi^0}(k^2)$, in Eq.(\ref{BS}) represents the usual $\pi^0$-meson
propagator:
\begin{equation}
  D_{\pi^0}(k^2) = \frac{1}{k^2-m_{\pi^0}^2} ~. \label{meson}
\end{equation}
\noindent At the same time, the (dressed) quark propagator is defined as  
\begin{equation}
 iS_q(x,x^\prime)\equiv \langle 0| T[\psi_q(x) \bar{\psi}_q (x^\prime)]|0 \rangle~,~q=u,d~,
\end{equation}
where $T$ is the time ordering operator. Using the standard quantum field theory procedure, i.e., 
expanding the fermion field operator, $\psi_q(x)$,  in a basis of states of the Dirac equation 
in a constant magnetic field, one obtains at zero temperature in coordinate space \cite{gus,kuz}:
\begin{equation}
 S_{q}(x,x')=e^{i\Phi_q (x,x')}\sum_{n=0}^{\infty}{S}_{q,n}(x-x')~,~q=u,d~, \label{prop1}
\end{equation}
\noindent the above propagator is given by the product of a gauge dependent factor $\Phi_q(x,x')$,
called Schwinger phase, times a translational invariant term. The Schwinger phase can be 
written in the Landau gauge as:
\begin{equation}
 \Phi_q(x,x') = Q_q \int_{x}^{x^\prime}dy_\mu A^\mu(y) ~,
\end{equation}
where the integral should be performed along a straight line connecting $x$ and $x^\prime$. Explicitly,  
one has:
\begin{equation}
 \Phi_q (x,x')=-\frac{Q_q B}{2}(x^{1}+x^{1'})(x^{2}-x^{2'}),
\end{equation}
\noindent The translation invariant part of the propagator is given by: 
\begin{multline}
S_{q,n}(Z)=\frac{\beta_{q}}{2\pi}exp\left(-\frac{\beta_{q}}{4}Z^{2}_{\perp} \right)
\int \frac{d^{2}p_{\parallel}}{(2\pi)^{2}}\frac{e^{(ip\cdot Z)_{\parallel}}}{p^{2}_{\parallel}-M^{2}-2
\beta_{q} n}  
\\ \times \left \{\left[\left(p\gamma\right)_{\parallel}+
M \right]\left[ \Pi_{-}L_{n}\left(\frac{\beta_{q}}{2}Z^{2}_{\perp}\right)+
\Pi_{+}L_{n-1}\left(\frac{\beta_{q}}{2}Z^{2}_{\perp}\right) \right] \right. \\
\left. +2in\frac{(Z\cdot \gamma_{\perp})}{Z_{\perp}^{2}}\left[L_{n}\left(\frac{\beta_{q}}{2}Z^{2}_{\perp}\right)-
L_{n-1}\left(\frac{\beta_{q}}{2}Z^{2}_{\perp}\right) \right]  \right \} ~.
\end{multline}
In the above expression $Z=x-x'$, $Z^{2}_{\parallel}=(Z^{2}_{0}-Z^{2}_{3})$, 
$Z^{2}_{\perp}=(Z^{2}_{1}+Z^{2}_{2})$,  $\Pi_{\pm}=\frac{1}{2}(\mathbb{I} \pm i\gamma^{1}\gamma^{2})$, 
$\beta_q=|Q_q|B$ and $L_{n}(x)$ represent the Laguerre polynomials.

From a direct comparison of  Eqs (\ref{BS}) and (\ref{meson}), one sees that under the 
present approximation the mass of the $\pi^0$-meson can be associated to the root
of the equation:
\begin{equation}
 1-2G\Pi_{ps}(k^{2})|_{k^2=m^2_{\pi^0}}=0~. \label{polemass}
\end{equation}
Substituting the quark propagator in the pseudoscalar polarization loop, Eq.(\ref{loop-ps}),
one obtains: 

\begin{multline}
 \frac{1}{i}\Pi_{ps}(k^{2})= -\sum_{q=u,d} ~\sum_{n,m=0}^{\infty}
 \int d^{4}(x~ -x^\prime) 
 e^{-ik\cdot (x-x^\prime)} \\
 \times Tr[i\gamma_{5}i S_{q,n}(x-x^\prime)i\gamma_{5}i S_{q,m}(x^\prime-x)]
 e^{i\Phi_q(x,x^\prime)}e^{i\Phi_q(x^\prime,x)} ~.
\end{multline}
\noindent Since the Schwinger phases cancel out, $e^{i\Phi_q(x,x')}e^{i\Phi_q(x',x)}=1$,  the 
Fourier transform 
can be applied to the last expression yielding:
\begin{multline}
 \frac{1}{i}\Pi_{ps}(k^{2})= -\sum_{q=u,d} ~\sum_{n,m=0}^{\infty}
 \int \frac{d^{4}p}{(2\pi)^4} \\
 \times Tr[\gamma_{5} S_{q,n}(p+\frac{k}{2})\gamma_{5} S_{q,m}(p-\frac{k}{2})] ~.
\end{multline}
%
%
Substituting the quark propagator, computing the traces, integrating over the Laguerre polynomials and performing a partial fraction 
decomposition, one obtains, after a rather long calculation, the expression:
\begin{multline}
 \frac{1}{i}\Pi_{ps}(k^{2})= \sum_{q=u,d}~\sum_{n=0}^{\infty} 2g_{n}\beta_{q} N_{c} \\  
 \times \left(\int \frac{d^{2}p_{\parallel}}{(2\pi)^{3}} \frac{1}{p_{\parallel}^{2}-M^{2}-2\beta_{q} n} \right. \\
 \left. -\int \frac{d^{2}p_{\parallel}}{(2\pi)^{3}}
 \frac{({k_{\parallel}^2}/2)}{(p_{\parallel}^{2}-M^{2}-2\beta_{q} n)((p+k)_{\parallel}^{2}-M^{2}-2\beta_{q} n)} 
  \right)~, \label{loop-2}
\end{multline}
where $g_{n}=2-\delta_{n0}$ , $p_{\parallel}=p_{0}-p_{3}$, and $k_{\parallel}=k_{0}-k_{3}$.

In the last expression the second integral can be rewritten using the Feynman integration trick 
and making an appropriate change of variables\cite {love,ryder}. Furthermore,  
the integral over
$p_0$ can be easily performed through a Wick rotation, resulting in the following expression for 
the polarization loop:
\begin{multline}
 \frac{1}{i}\Pi_{ps}(k^{2})= - \sum_{q=u,d}~\sum_{n=0}^{\infty} 2g_{n}\beta_{q} N_{c}\\  
 \times \left( i \int \frac{dp_3}{(2\pi)^3}  \frac{\pi}{ \sqrt{p_{3}^{2}+M^{2}+2\beta_{q} n} } +  
 \frac{k_{\parallel}^{2}}{2(2\pi)^{3}}I_{q,n}(k^{2}) \right)~, \label{loop-3}
\end{multline}
where the integral $I_{q,n}(k^{2})$ is defined as:

\begin{equation}
 I_{q,n}(k^{2})=\frac{i\pi}{2} \int_{0}^{1}dx\int_{-\infty}^{\infty} dp_3
 \frac{1}{[p^{2}_{3}+\overline{M}^{2}(k_{\parallel})+2\beta_{q} n]^{3/2}} ~, \label{intloop}
\end{equation}

\noindent and  $\overline{M}^{2}(k_{\parallel})=M^{2}-x(1-x)(k_{\parallel}^{2})$ \cite{love}.
 
The polarization loop expression, Eq.(\ref{loop-3}), can be further simplified if we consider 
the mass gap expression which be obtained from Eq.(\ref{gap}) as follows:
\begin{eqnarray}
 M&=&m-2G  \langle \bar{\psi} \psi \rangle~ \nonumber \\
 &=&~m+2G\lim_{t^\prime \to t^+} \lim_{x^\prime \to x} \sum_{q=u,d} Tr[ iS_q(x,x^\prime) ]~.
\end{eqnarray}
The last equation can be calculated using the quark propagator, Eq.(\ref{prop1}), and 
noticing that the Schwinger phase corresponds to $\Phi_q(x,x)=1$. Performing such calculations one then finds
the usual gap equation \cite{nosso1}: 
\begin{multline}
 \frac{M-m}{2MG}=\\\sum_{q=u,d}~\sum_{n=0}^{\infty} 2 g_{n}\beta_{q} N_{c}
 \int_{-\infty}^{\infty} \frac{dp_3}{(2\pi)^3}  \frac{ \pi}{ \sqrt{p_{3}^{2}+M^{2}+2\beta_{q} n} } ~,
\end{multline}
where $N_c$=3. Thus, substituting the last expression in Eq.(\ref{loop-3}) the loop polarization becomes: 
\begin{multline}
 \frac{1}{i}\Pi_{ps}(k^{2})= -i\left(\frac{M-m}{2MG}\right) \\ 
- \sum_{q=u,d} \beta_{q} 
 N_{c}\frac{k_{\parallel}^{2}}{(2\pi)^{3}}\sum_{n=0}^{\infty}g_{n} I_{q,n}(k^{2})~. \label{polar}
\end{multline}
Therefore, from Eq.(\ref{polemass}), the $\pi^0$ mass can be written as:
%
 \begin{equation}
 m_{{\pi}^0}^{2}(B)=-\frac{m}{M(B)}
 \frac{(2\pi)^{3}}{\displaystyle{\sum_{q=u,d}  i2G\beta_{q} N_{c}\sum_{n=0}^{\infty}
 g_{n}I_{q,n}(m_{{\pi}^0}^{2})} }~. \label{mpion}
\end{equation}
In  the present work, the divergent integrals such as $I_{q,n}$ are  regularized by a sharp  noncovariant
cutoff, $\Lambda$, within the MFIR scheme. This method, which has been reported in Ref. \cite {nosso1}, follows 
the steps of the dimensional regularization prescription of QCD, performing a sum
over all Landau levels in the vacuum term. This allows to isolate the divergencies into a term
that has the form of the zero magnetic field vacuum energy and that can be regularized in
the standard fashion. We point out that originally such a method was proposed in Ref. \cite {klimenko}  but
in terms of the proper-time formulation. Recently, the MFIR has been successfully used in the case of magnetized 
color superconducting cold matter \cite{norberto} where its advantages, such as the avoidance of unphysical 
oscillations, are fully discussed.  

Then, following the  techniques illustrated in  Refs \cite{nosso1,klimenko} which consider the Hurwitz-Zeta function integral 
representation, the sum over Landau levels  can be performed and the regularized integral  reads (see appendix):
\begin{multline}
 \sum_{n=0}^{\infty}g_{n}I_{q,n}(k^{2}=m^2_{{\pi}^0})=\frac{i\pi}{\beta_{q}}\int^{1}_{0}dx \\ 
 \times \left[ -\psi\left(\frac{\overline{M}^{2}(m_{\pi^0})}{2\beta_{q}}+1 \right)  +
\left(\frac{\beta_{q}}{\overline{M}^{2}(m_{\pi^0})} \right)  + 
 \ln\left({\frac{\overline{M}^{2}(m_{\pi^0})}{2\beta_{q}}}\right) \right. \\  
\left. -2\left(\frac{\Lambda}{\sqrt{\Lambda^{2}+\overline{M}^{2}(m_{\pi^0})}}-
\sinh^{-1}{\frac{\Lambda}{\overline{M}(m_{\pi^0})}}\right) \right]  ~, \label{int00}
\end{multline}
\noindent where ${\psi}$ is the digamma function. 
\noindent The $\sigma$-meson mass, $m_\sigma$, can be obtained in a completely analogous fashion
by calculating the scalar polarization loop, Eq.(\ref{loop-sc}), yielding:  
\begin{equation}
m_{\sigma}^{2}(B)=4M^{2}(B)+m_{\pi^{0}}^{2}(B)\,.
\end{equation}
%
%
%
%
\noindent We next discuss the calculation of the pion decay constant, $f_{{\pi}^0}$, which implies 
the evaluation of   following matrix element\cite{kleva}:

\begin{equation}
 \left \langle 0\left | J_{5_{\mu}}^{i} \right | \pi^{i}\right \rangle~,
\end{equation}

\noindent  that is equivalent to:

\begin{multline}
 ik_{\mu}f_{\pi^{0}}\delta_{ij}= \sum_{q=u,d}\\ 
 -\int\frac{d^{4}p}{(2\pi)^{4}}
 Tr\left[i\gamma_{\mu}\gamma_{5}\frac{\tau^{i}}{2}iS_q(p+\frac{1}{2}k)ig_{\pi qq}\gamma_{5}\tau^{j}
 iS_q(p-\frac{1}{2}k)\right] ~.
\end{multline}
\noindent After evaluating the trace one obtains:
\begin{equation}
 f_{\pi^{0}}^{2}(B)=-i\sum_{u,d}\frac{\beta_{q}}{(2\pi)^{3}}N_{c}M^{2}
 \sum_{n=0}^{\infty} g_{n}I_{q,n}(0) ~, \label{fpion}
\end{equation}

\noindent where $I_{q,n}(0) \approx I_{q,n}(m_{\pi}^{2})$ is the integral 
given in Eq.(\ref{int00}). Multiplying the pion mass, Eq.(\ref{mpion}), by the pion decay 
constant, Eq.(\ref{fpion}), one obtains:
\begin{equation}
 m_{\pi^0}^2(B)f_{\pi^{0}}^{2}(B)= \frac{m~M(B)}{2G}~.
\end{equation}
\noindent Now,  Eq.(\ref{gap}) can be used in order to eliminate the coupling constant, $G$, of the 
last equation leading to the Gell-Mann--Oakes--Renner (GOR) relation in a magnetic medium

\begin{equation}
 m_{\pi^{0}}^{2}(B) f_{\pi^{0}}^{2}(B)=m \left \langle \overline{\psi}\psi\right\rangle(B).
 \label{gor}
\end{equation}

\section{Numerical Results}

\begin{figure}[!h]
\begin{tabular}{ccc}
\includegraphics[width=8.cm]{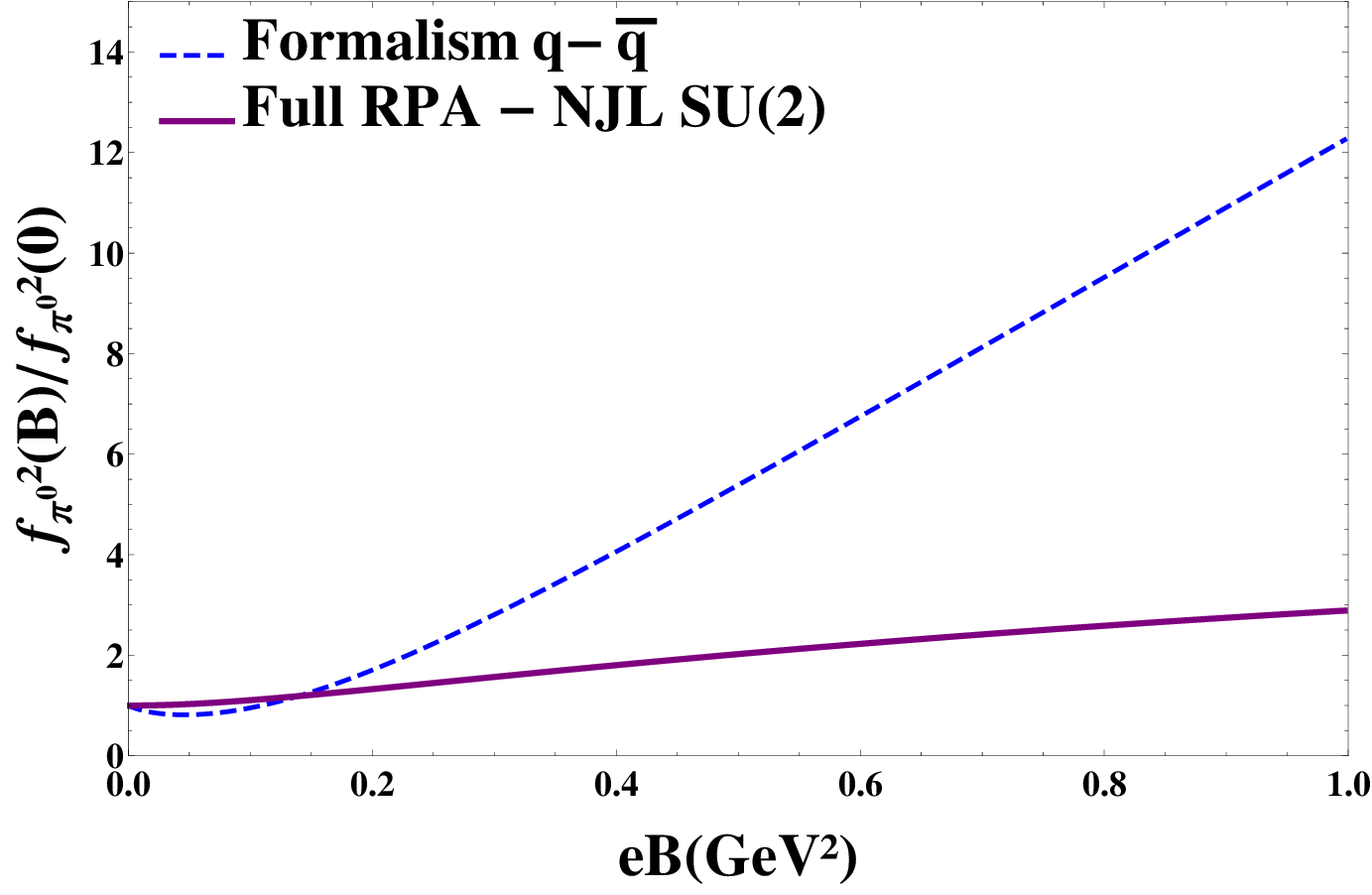}\\

\end{tabular}
\caption{Normalized pion decay constant $f_{\pi^{0}}^{2}(B)/f_{\pi^{0}}^{2}(0)$  evaluated within the 
full RPA and the ${\bar q} q$ formalism of Ref. \cite {simonov} up to $eB=1\, {\rm GeV}^{2}$. }
\label{fpisimonov}
\end{figure}

\begin{figure}[!h]
\begin{tabular}{ccc}
\includegraphics[width=8.cm]{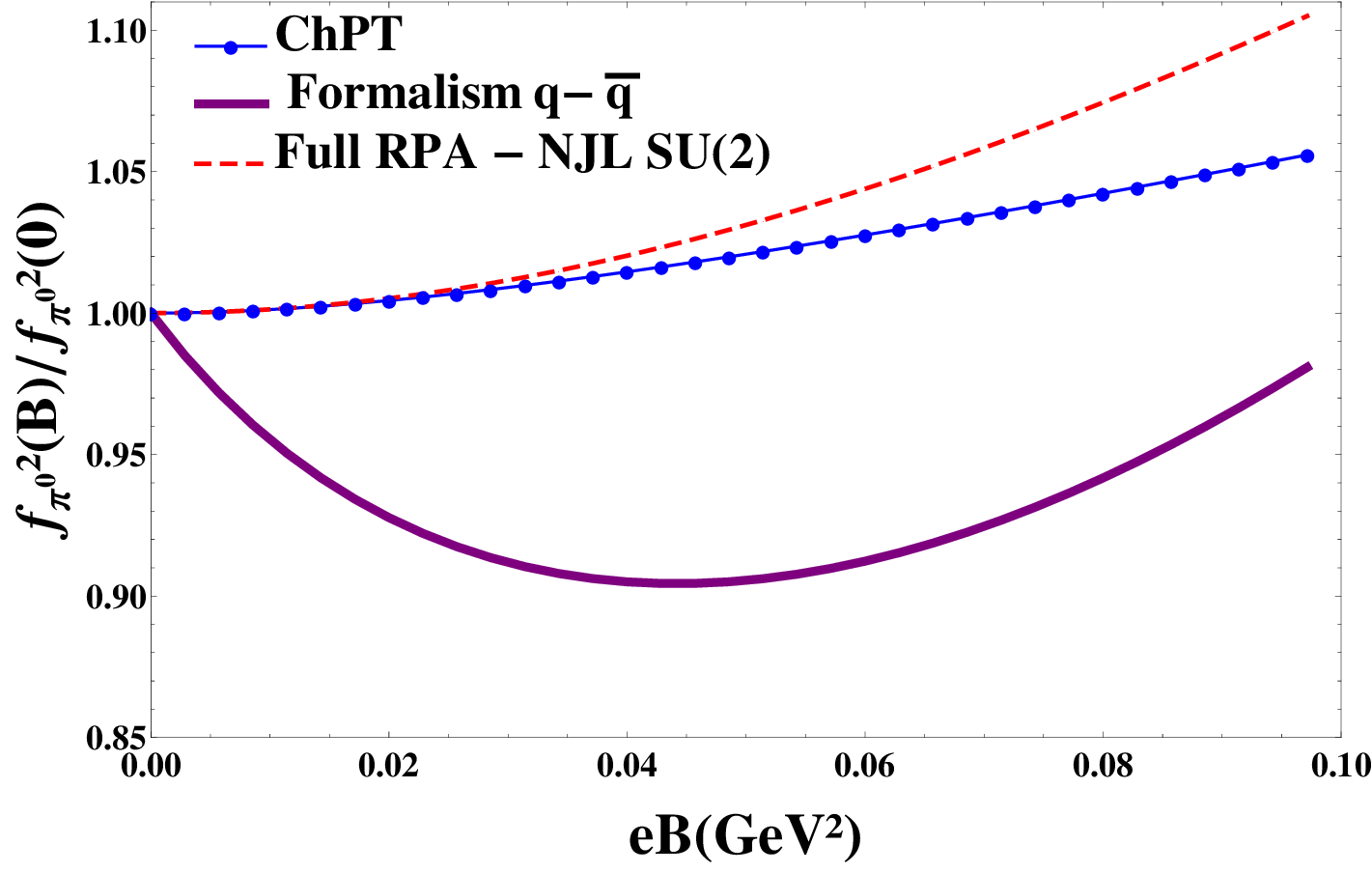}\\

\end{tabular}
\caption{Normalized pion decay constant $f_{\pi^{0}}^{2}(B)/f_{\pi^{0}}^{2}(0)$  evaluated within the 
full RPA, the ${\bar q} q$ formalism  \cite {simonov}, and ChPT \cite {andersen} up to $eB=0.1\,{\rm GeV}^{2}$.}
\label{fpiandersen}
\end{figure}

\begin{figure}[!h]
\begin{tabular}{ccc}
\includegraphics[width=8.cm]{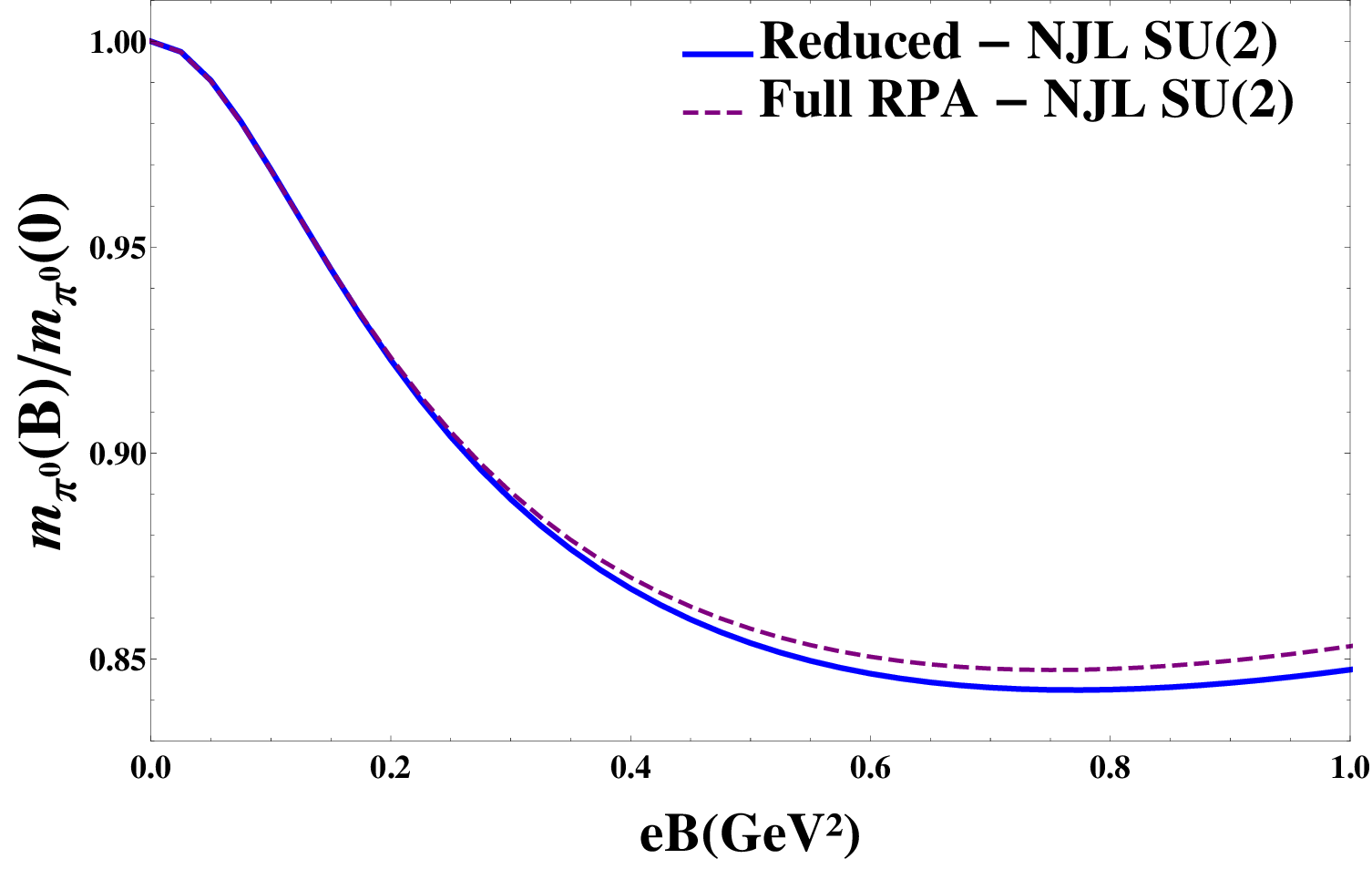}\\

\end{tabular}
\caption{Normalized neutral meson mass,  $m_{\pi^{0}}(B)/m_{\pi^{0}}(0)$, evaluated with the RPA using 
the complete polarization integral as well as the approximation $I(m_{\pi}^2)\approx I(0)$.}
\label{mpiFullvs0}
\end{figure}

\begin{figure}[!h]
\begin{tabular}{ccc}
\includegraphics[width=8.cm]{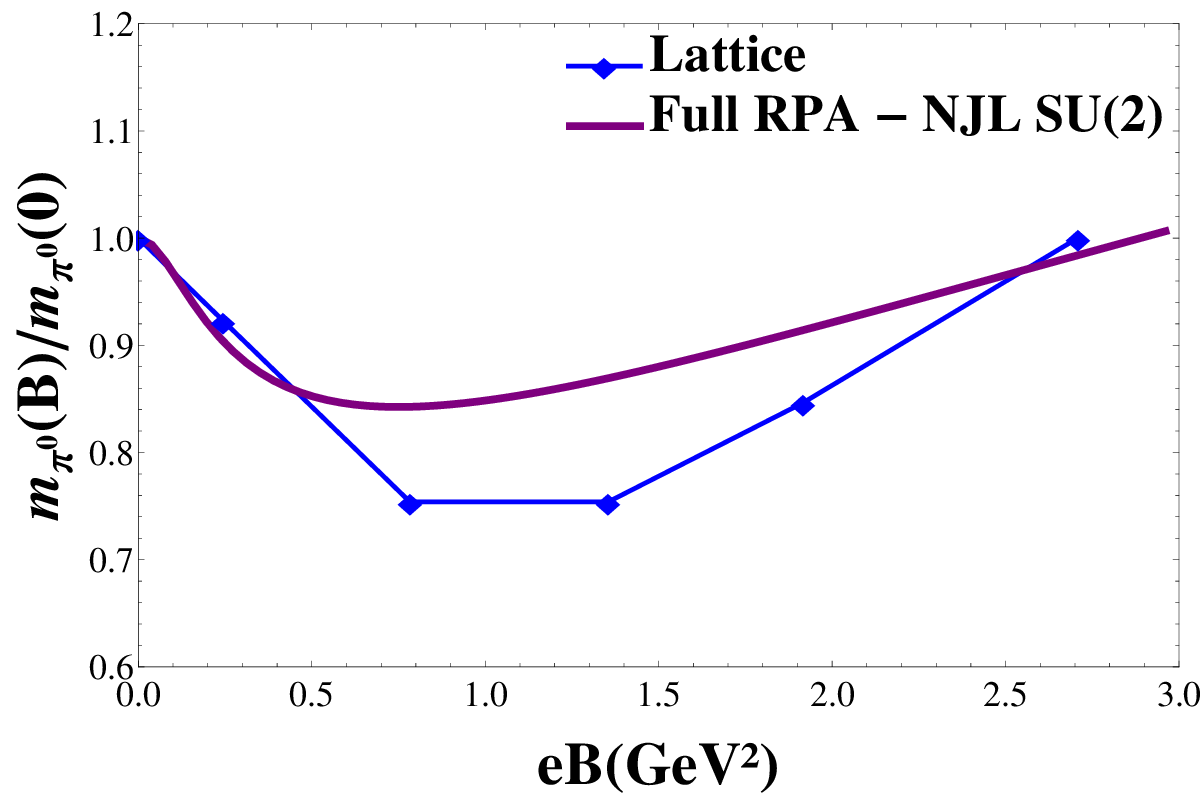}\\

\end{tabular}
\caption{Normalized neutral meson mass,  $m_{\pi^{0}}(B)/m_{\pi^{0}}(0)$, evaluated with the full RPA
and LQCD \cite {hidaka}. Within the latter, the lines are shown just to guide the eye. }
\label{massLQCD}
\end{figure}

\begin{figure}[!h]
\begin{tabular}{ccc}
\includegraphics[width=8.cm]{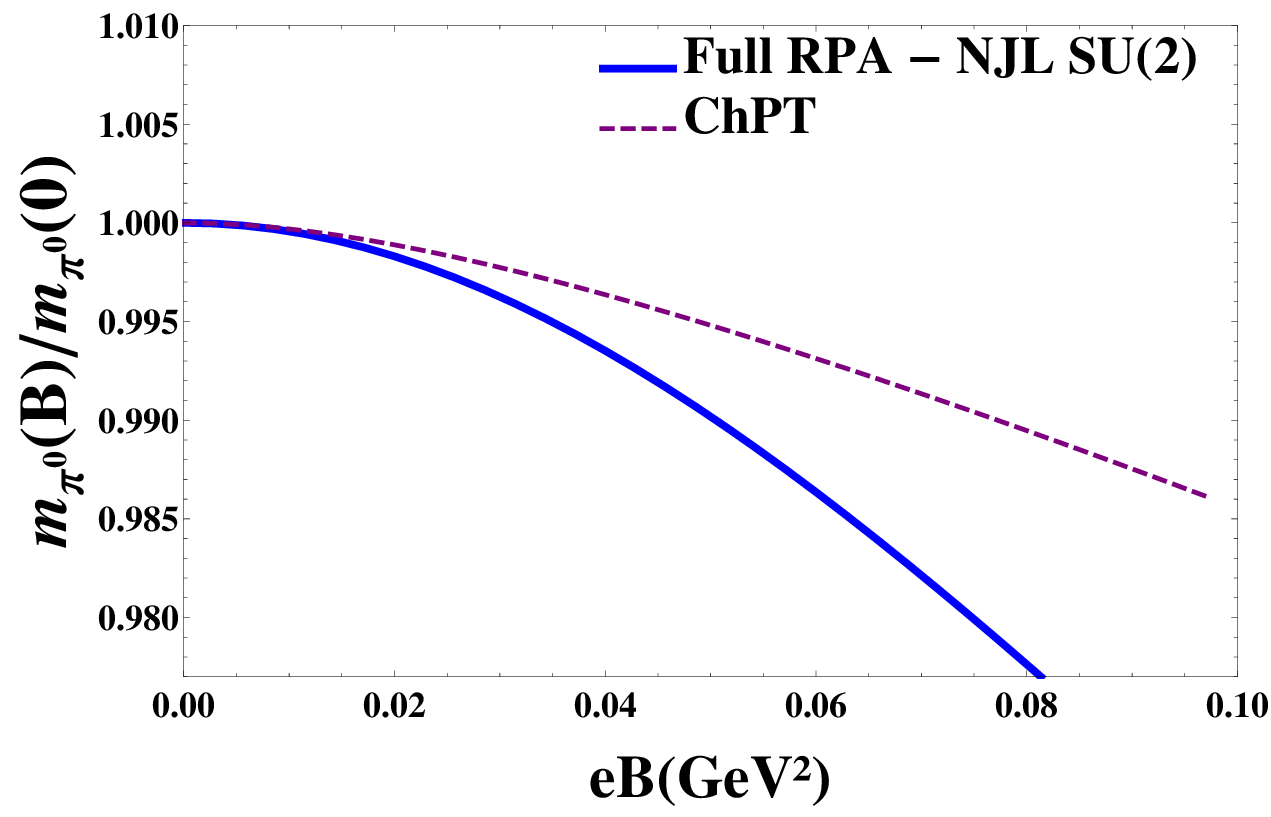}\\

\end{tabular}
\caption{Normalized neutral meson mass,  $m_{\pi^{0}}(B)/m_{\pi^{0}}(0)$, evaluated with the full RPA
and ChPT \cite {andersen}. }
\label{massChPT}
\end{figure}

\begin{figure}[!h]
\begin{tabular}{ccc}
\includegraphics[width=8.cm]{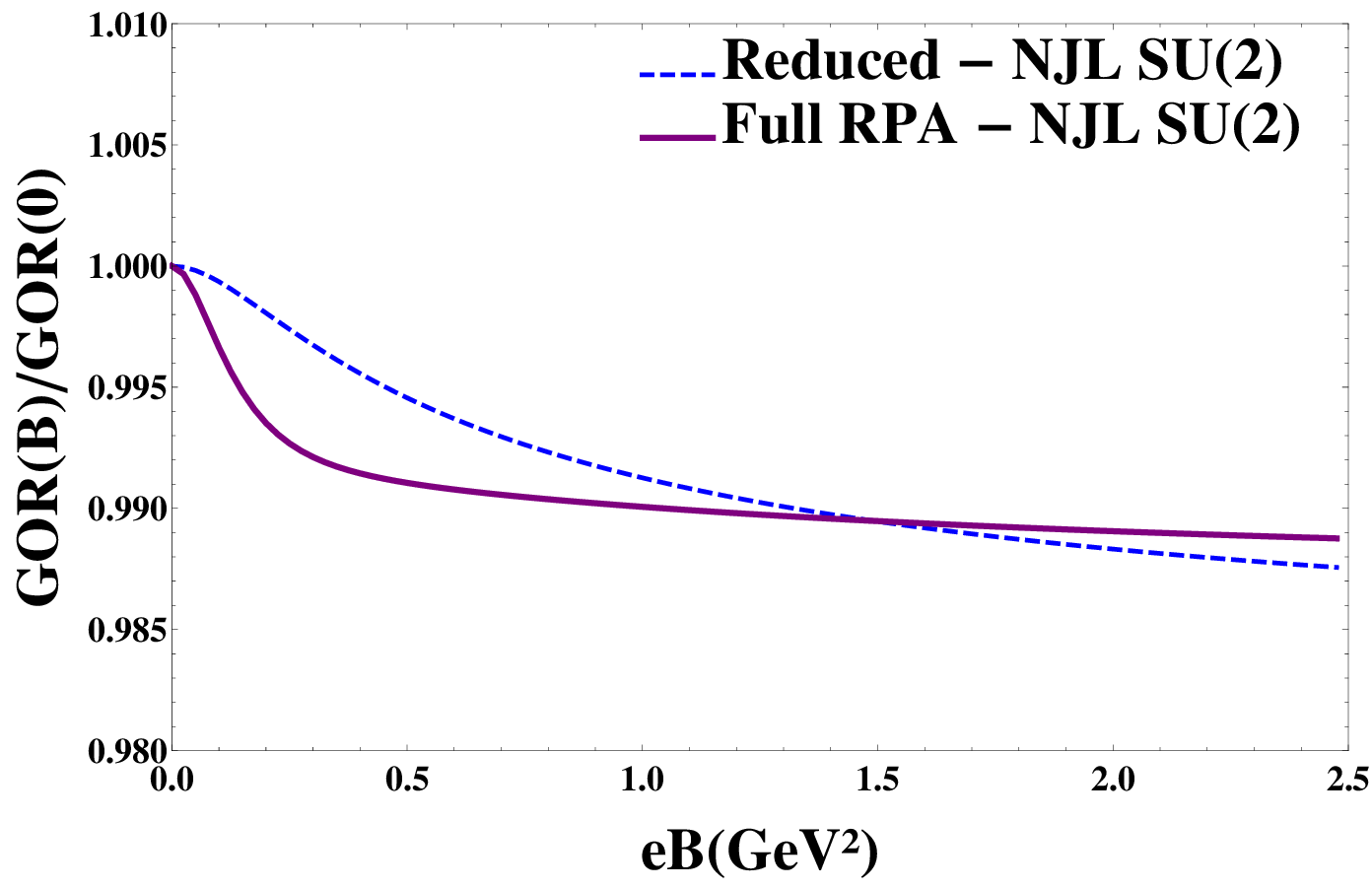}\\

\end{tabular}
\caption{Normalized Gell-Mann-Oakes-Renner relation, ${\rm GOR}(B)/{\rm GOR}(0)$, as given by
Eq. (\ref {gor}), evaluated with the RPA using the complete polarization integral as well as the 
approximation $I(m_{\pi}^2)\approx I(0)$. }
\label{gorf}
\end{figure}

Let us now perform numerical evaluations in order to compare our results with those from other
applications. In this work we choose the following parametrization 
set \cite {buballa} : $ \Lambda=664.3\,{\rm MeV}, m=5.0\, {\rm MeV}$, and $G \Lambda^{2}=2.06$ which
reproduce $f_\pi^0 = 92.4 \, {\rm MeV}$, $m_\pi^0=  135.0\, {\rm MeV}$, and 
$\langle {\bar q} q \rangle^{1/3}= -250.8 \, {\rm MeV}$. Fig \ref {fpisimonov} compares our result 
for $f_{\pi^{0}}$ with the one obtained, with the $q {\bar q}$ formalism in Ref. \cite {simonov} 
for $eB$ up to $1 \, {\rm GeV}^2$. For very weak field values the figure shows that both approximations predict an 
opposite behavior with the $q {\bar q}$ formalism predicting an initial decrease of $f_\pi^0$.  Then, for fields higher than about
$0.05 \,{\rm GeV}^2$, both approximations predict that $f_\pi^0$ increases with $B$  with the full RPA predicting a less dramatic increase. 
In order to  settle the   qualitative disagreement occuring  within the weak field range we compare, 
in Fig \ref {fpiandersen}, both results with the  ChPT  predictions for low $eB$ values \cite {andersen}. 
The figure shows a good agreement between the full RPA and ChPT while the $q {\bar q}$ formalism 
predicts an opposite behavior. We remark that in the  evalution of $f_{\pi^0}$ one always considers 
the polarization integral $I(k^2=0)$ so that our result for this quantity agrees with the ones 
obtained in early evaluations \cite {kleva}. Next, in Fig. \ref {mpiFullvs0}, we compare the results 
generated by the RPA with, and without, the further approximation $I(k^2) \approx I(0)$ for the 
mass pole. A visible, although very small, difference is observed at intermediate $eB$ values 
while at high field intensity, when only the LLL contribute, the  results coincide. It is important 
to note that in Ref. \cite {kleva}, where the RPA with $I(k^2=0)$ has been used, the pionic mass 
seems to be almost insensitive to $B$ but that is due to the fact of using an inconsistent  regularization 
procedure in the purely magnetic sector (in that work all the magnetic dependence  enters via the 
effective quark mass only). In Fig \ref {massLQCD} our full RPA results are compared with the 
LQCD predictions of Ref. \cite {hidaka} showing a 
rather good agreement up to about $0.5 \, {\rm GeV}^2$ whereas for intermediate field values 
the maximum difference is not higher than about $15 \, \%$ at intermediate values dropping 
again for high values ($eB \gtrsim 2 \, {\rm GeV}^2$).  For lower field 
strength values, our predictions can also be compared with the ChPT available 
in Ref. \cite {andersen} as Fig. \ref {massChPT} shows. The agreement is also good for this range 
where LQCD results are not available. Now, having evaluated $f_{\pi^0}$, $m_{\pi^0}$ as well as the 
quark condensate, $\langle {\bar \psi} \psi \rangle$, one may investigate how the GOR relation, given 
by Eq. (\ref {gor}), holds in the presence of a magnetic field. Fig \ref {gorf} shows the behavior of 
this quantity within the RPA formalism using the complete polarization integral 
and the approximated one ($I(k^2) \approx I(0)$. The figure shows that after dropping from the $B=0$ 
value by about only $1\%$  the full RPA result remains quite stable when $B$ increases in agreement with
the analytical predictions by Agasian and Shushpanov \cite {agasian}. On the other hand, the results obtained 
with the  approximated integral appear to be 
less stable at high $B$ values. 

\begin{figure}[!h]
\begin{tabular}{ccc}
\includegraphics[width=8.cm]{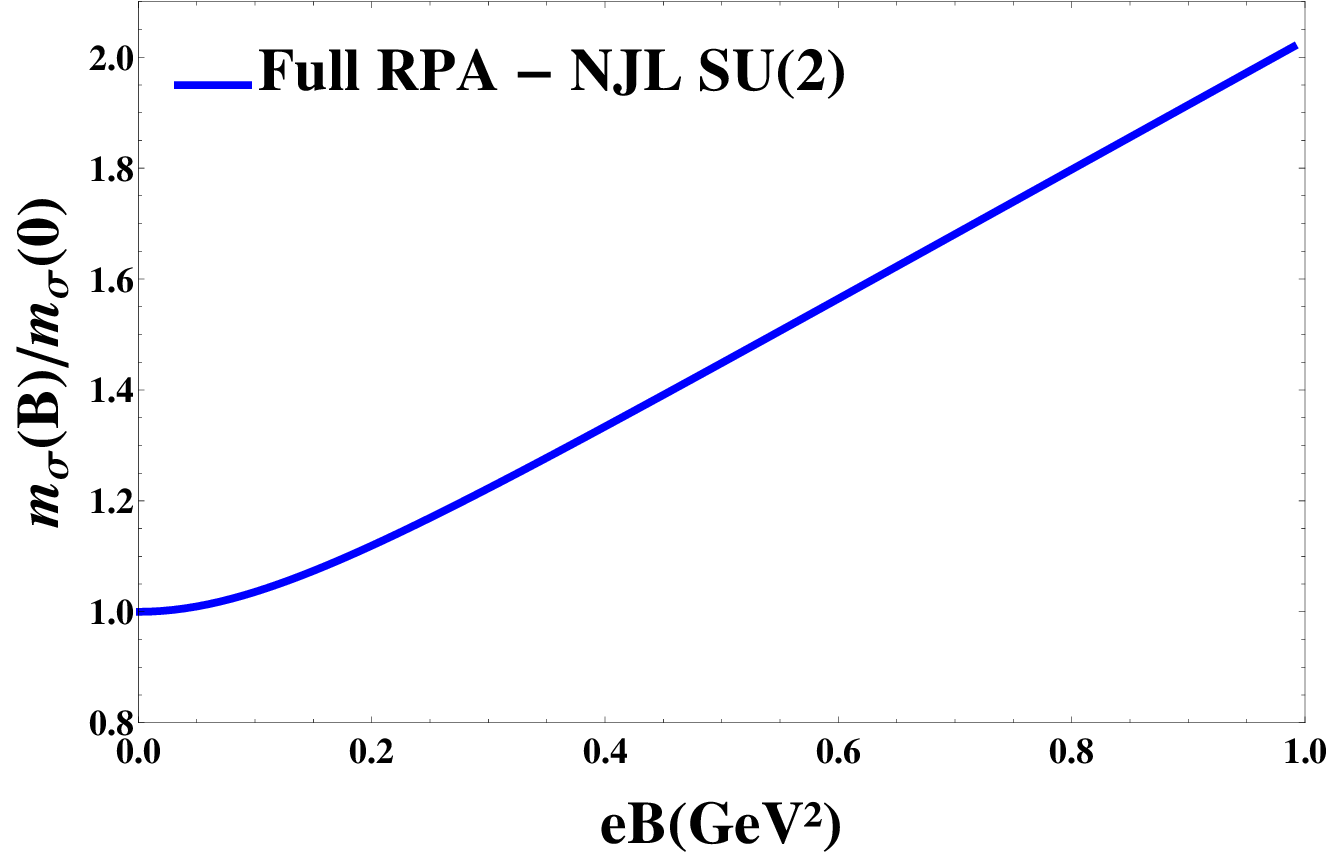}\\

\end{tabular}
\caption{ Normalized $\sigma$ meson mass, $m_\sigma(B)/m_\sigma(0)$, evaluated with the full RPA 
formalism developed in this work.}
\label{Msigma}
\end{figure}

\begin{figure}[!h]
\begin{tabular}{ccc}
\includegraphics[width=8.cm]{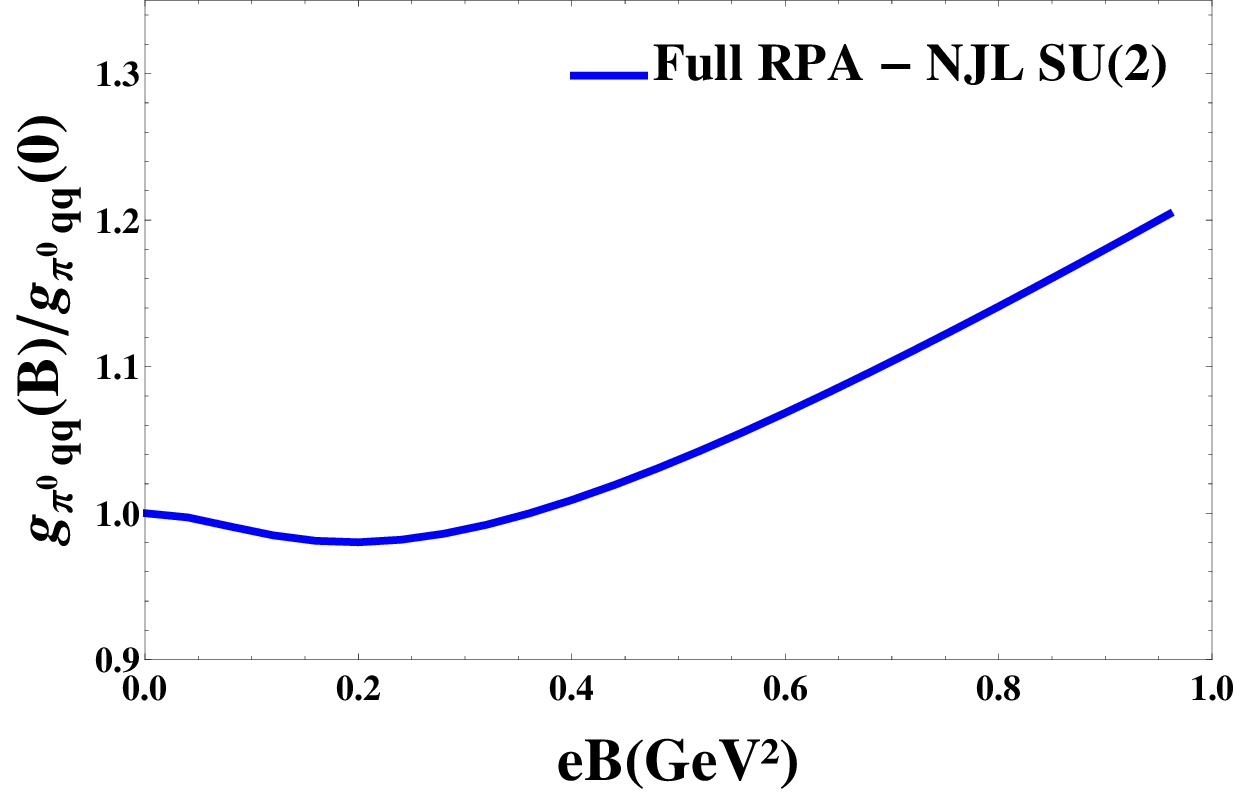}\\

\end{tabular}
\caption{ Normalized meson-quark coupling, $g_{\pi^{0} qq}(B)/g_{\pi^{0}}qq$, evaluated with the
full RPA formalism.}
\label{gpiqq}
\end{figure}

Concerning  the  $\sigma$ meson mass our results, shown in Fig. \ref {Msigma}, confirm the steady increase 
observed in Ref. \cite {kleva}.  Finally, 
Fig. \ref {gpiqq} shows how the meson-quark coupling, $g_{\pi^0 qq}$, slightly decreases up 
to $eB \approx 0.2 \, {\rm GeV}^2$ and then steadily increases at higher field values. 

\section{Conclusions}

We have used the RPA without any further approximations  to evaluate quantities related to 
magnetized neutral mesons. In this way, the momentum dependence of the integral representing 
the polarization function has been fully taken into account in order to render more accurate
results. After performing   the summation over the Landau levels we have obtained  results  
regularized in a consistent way, through the use of a magnetic field independent regularization 
scheme, which  is a  welcome feature since    the polarization function
contributes to physical quantities such as the mass pole and decay constants. Our numerical 
results for these observables are in good agreement with other powerful methods such as ChPT 
and LQCD. It is also  important to remark that, within magnetized hadronic matter, the more 
cumbersome evaluations related to mesonic  contributions have received less attention than the 
contributions due to quarks which stem from the uncomplicated zero external momentum Green's 
functions such as the ones considered in the evaluation of quantities like the (mean field) 
equation of state or quark condensates. 
Therefore, by  presenting the complete evaluation of the magnetized polarization tensor, our  
work provides a framework for investigations where this quantity plays a crucial role as, e.g.,  
the equation of state with pionic terms \cite {fukushima} and  shear viscosity \cite {lang} among others. 
The method can also be readily extended to accommodate charged mesons and thermal effects.

\appendix
\section{Regularization procedure}
In this appendix we  present some technical details in order to obtain our main 
analytical result, i.e.,  Eq.(\ref{int00}). This procedure is based on the magnetic field independent 
regularization scheme \cite {nosso1,klimenko}. We start with: 
\begin{multline}
 \sum_{n=0}^{\infty}g_{n}I_{q,n}(k^{2})=\\
 \sum_{n=0}^{\infty}g_{n}\int d^{2}p_{\parallel}\frac{1}{[p^{2}_{\parallel}-M^{2}-2\beta_{q} n]
 [(p+k)_{\parallel}^{2}-M^{2}-2\beta_{q} n]}\, . \label{app1}
\end{multline}
The last integral can be rewritten using Feynman's parametrization\cite{love,ryder}
\begin{equation*}
\frac{1}{AB}=\int_0^1 dx \frac{1}{\left[ xA+(1-x)B \right]^2 } \;,
\end{equation*}
making a change of variable ($ p_\parallel \rightarrow p_\parallel + x~k_\parallel $) and defining 
$ \overline{M}^{2}(k_{\parallel})=M^{2}-x(1-x)k_{\parallel}^{2}$. Thus,  eq.(\ref{app1}), becomes: 
\begin{multline}
\sum_{n=0}^{\infty}g_{n}I_{q,n}(k^{2})= \\
\int_{0}^{1}dx \sum_{n=0}^{\infty}g_{n}\int d^{2}p_{\parallel}\frac{1}{[p^{2}_{\parallel}-\overline{M}(k_\parallel)^{2}-
2\beta_{q} n]^2}
\\ = \frac{i\pi}{2(2\beta_{q})^{3/2}} \int_{0}^{1}dx
\sum_{n=0}^{\infty}g_{n}\int_{-\infty}^{\infty} dp_{3}\frac{1}{\left[\frac{p^{2}_{3}+\overline{M}^{2}(k_{\parallel})^{2}}{2\beta_{q} }+
n \right]^{3/2}} \;,  \label{app2}
\end{multline}
where in the last expression the $p_{0}$-integral was performed through a Wick rotation and for convenience 
$2(\beta_{q})^{3/2}$ was factored out. Now, the summation over $n$ in eq.(\ref{app2}) can be done by using the  
Riemann-Hurwitz zeta function,
\begin{equation}
 \zeta(z,x)=\sum_{n=0}^{\infty}\frac{1}{(x+n)^{z}}\;,
\end{equation}
and its integral representation\cite{grad}
\begin{multline}
 \int_{0}^{\infty}dyy^{z-1}\exp[-\kappa y]\coth(\alpha y)=\\ 
 \Gamma[z]\left(2^{1-z}\alpha^{-z}\zeta(z,\frac{\kappa}{2\alpha})-\kappa^{-z} \right) \;, \label{zeta}
\end{multline}
with the  identification:
\begin{equation}
 \alpha=|Q_{q}|B \equiv \beta_{q}, \quad  \kappa=\overline{M}^{2}(k_{\parallel})+p_{3}^{2}, 
 \quad z=\frac{3}{2}\;.
\end{equation}
After some algebraic manipulations one obtains
\begin{equation}
 I_{t} \equiv \sum_{n=0}^{\infty}g_{n} I_{q,n}=i\pi\int_{0}^{1}dx\int_0^{\infty} dy 
 \exp(-\overline{M}^{2}y)\coth(\beta_{q} y)
\;. \label{itdef}
\end{equation}
Expanding $\coth(\beta_{q}y)$ as a power series in $\beta_{q}y$, we find for $I_t$  :
\begin{equation}
 I_{t}=i\pi \int_{0}^{\infty}dx\int_0^{\infty} dy \exp(-\overline{M}^{2} y)\left(\frac{1}{\beta_{q} y}+
 \frac{\beta_{q} y}{3}+...\right)\;. \label{It}
\end{equation}
Now, for convenience we define:
 \begin{equation}
  I_{a}=i\pi\int^{1}_{0}dx\int_0^{\infty} dy \exp(-\overline{M}^{2} y)\frac{1}{\beta_{q} y}\;. \label{Ia}
 \end{equation}
 Note that for all the observables of interest, e.g., eqs.(\ref{polar},\ref{mpion},\ref{fpion}), $I_t$ 
always appears as the product $\beta_q\times I_t$. Hence, we split $I_t$  
in two terms,
 \begin{equation}
  I_{t}=(I_{t}-I_{a})+I_{a}\;,
 \end{equation}
the first term in parentheses results to be a $B$-dependent finite contribution for the observables and the second  
a $B$-independent infinity one, which needs to be regularized.
Let us now evaluate explicitly $I_{t}-I_{a}$, which is given by:
\begin{equation}
I_{t}-I_{a}=i\pi\int_{0}^{1}dx\int dy\exp(-\overline{M}^{2}y)\left(\coth(\beta_{q} y)-
\frac{1}{\beta_{q} y}\right)\,.
\end{equation}
We can be rearrange the above expression  using the gamma function integral representation\cite{grad} ,  
\begin{equation}
 \frac{\Gamma[z+1]}{{\beta}^{z+1}}=\int_0^{\infty}dy~y^z {e}^{-\beta y} \;, \label{gammafu}
 \end{equation}
and again the zeta function integral representation, eq.(\ref{zeta}), obtaining:
\begin{multline}
I_{t}-I_{a}= \lim_{\epsilon\rightarrow 0}  \frac{i\pi}{{\beta}_q}\int_{0}^{1} dx  \\
\left[ \Gamma[1+\epsilon]\left(2^{-\epsilon}\zeta(1+\epsilon,x_q)-
(2 x_q)^{-\epsilon-1} \right) - \frac{\Gamma(\epsilon)}{(2x_q)^{\epsilon}} \right] \;, \label{loopreg}
\end{multline}
where $x_q={\overline{M}^2}/({2{\beta}_q})$. The limit $\epsilon \rightarrow 0$ may be obtained 
using $a^{-\epsilon}\approx 1-\epsilon\ln{a}$ and the following properties\cite{grad}:
\begin{equation}
 \Gamma[\epsilon]=\frac{1}{\epsilon}-\gamma_{E}+O(\epsilon) ~~,~~ \Gamma[\epsilon+1]=\epsilon\Gamma(\epsilon)\; ,
 \end{equation}
\begin{equation}
 \zeta(1+\epsilon,a)=\lim_{\epsilon\rightarrow 0} \frac{1}{\epsilon}-\psi(a)\;,
\end{equation}
where in the last expression $\psi(x)$ is the digamma function\cite{grad,whittaker} and $\gamma_E$ the Euler-Mascheroni 
constant. Thus, one obtains:
\begin{equation}
  I_{t}-I_{a}=\frac{i\pi}{\beta_{q}}\int_{0}^{1}dx\left(-\Psi(x_q)-\ln 2 -(2x_{q})^{-1}+\ln 2x_{q}\right)\;.
\end{equation}
Finally, we have just to evaluate $I_{a}$, eq.(\ref{Ia}). Starting from the gamma function 
representation, eq.(\ref{gammafu}), one obtains:
\begin{multline}
\begin{aligned}
 & I_{a}=\frac{i\pi}{\beta_{q}}\lim_{\epsilon \rightarrow 0} \int_{0}^{1}dx\int_0^{\infty} dy
 \exp(-\overline{M}^{2} y)
 ~y^{-1+\epsilon}\\
 & =\frac{i\pi}{\beta_{q}}\lim_{\epsilon \rightarrow 0}\int_{0}^{1}dx
 \frac{\Gamma[\epsilon]}{(\overline{M}^{2})^{\epsilon}}\;, \label{iaa}
 \end{aligned}
\end{multline}
Here we use the integral representation of the Beta function\cite{grad} 
\begin{equation}
\int_{0}^{\infty}dxx^{\mu-1}(1+x^{2})^{\nu-1}=\frac{1}{2}B\left(\frac{\mu}{2},1-\nu-\frac{\mu}{2} \right)\;. 
\end{equation}
and the property
\begin{equation}
 B(x,y)=\frac{\Gamma[x]\Gamma[y]}{\Gamma[x+y]}\,,
\end{equation}
to write:
\begin{equation}
 \int_{0}^{\infty}\frac{p^{2}dp}{(p^{2}+\overline{M}^2)^{3/2}}= \lim_{\epsilon\rightarrow 0}
 \frac{1}{2(\overline{M}^{2})^{\epsilon}}
 B(\frac{3}{2},\epsilon)=
 \lim_{\epsilon\rightarrow 0}\frac{ \Gamma[\epsilon]}{2(\overline{M}^{2})^{\epsilon}}\;,
\end{equation}

Thus, comparing this result with eq.(\ref{iaa}), we conclude that the usual vacuum term of the NJL model 
without external magnetic field is recovered and again a regularization is necessary. We choose to use the 
three-momentum noncovariant cutoff scheme. In this case, the $I_{a}$ integral can be written as 
\begin{equation}
 I_{a}=-2\frac{i\pi}{\beta_{q}}\int_{0}^{1}dx\left[\frac{\Lambda}{\sqrt{\Lambda^{2}+
 \overline{M}^{2}}}-\sinh^{-1}\frac{\Lambda}{\overline{M}} \right] \,.
\end{equation}

Finally, considering the above results one obtains

\begin{multline}
I_{t}=\frac{i\pi}{\beta_{q}}\int_{0}^{1}dx\left[-\psi(x_{q}+1)+\frac{1}{2}x_{q}^{-1}+
\ln\left(x_{q}\right)-\right.\\ 
\left. 2\left(\frac{\Lambda}{\sqrt{\Lambda^{2}+\overline{M}^{2}}}-\sinh^{-1}
\frac{\Lambda}{\overline{M}} \right) \right]
\;,\\
\end{multline}
which leads to Eq. (\ref{int00}).

\section*{Acknowledgments}

This work was  partially supported by Conselho Nacional de Desenvolvimento 
Cient\'{\i}fico e Tecnol\'{o}gico (CNPq) and Coordenação de Aperfeiçoamento de Pessoal de Nível Superior (CAPES).

\end{document}